\documentclass[aps,prapplied,reprint,superscriptaddress]{revtex4-2}

\usepackage{graphicx}  
\usepackage{dcolumn}   
\usepackage{bm}        
\usepackage{amssymb}   
\usepackage{amsmath}
\usepackage{xcolor}
\usepackage{hyperref}

\usepackage{amsmath}
\usepackage{amsfonts}
\hypersetup{colorlinks=true, linkcolor=blue, citecolor=blue, urlcolor=blue}

\long\def\comment#1{}
\begin{document}


\title{Magnetic wideband VHF localized field probe using magnon polaritons}



\author{G. Soares}\email[Corresponding author: ]{gabrsoares@proton.com} \affiliation{INFN, Laboratori Nazionali di Legnaro, Legnaro, Padova, Italy - Current address Grenoble}
\author{N. Crescini} \affiliation{Fondazione Bruno Kessler (FBK), I-38123, Trento, Italy}
\author{G.~Carugno} \affiliation{INFN, Sezione di Padova, Padova, Italy} 
\author{G.~Ruoso}\email[Corresponding author: ]{Giuseppe.Ruoso@lnl.infn.it}\affiliation{INFN, Laboratori Nazionali di Legnaro, Legnaro, Padova, Italy}

\collaboration{SHRIMP collaboration}


\date{\today}

\begin{abstract}
We present here an optimisation and demonstration of a wide band instrument capable of measuring localised and directionally alternated magnetic fields below pT in the very high frequency (VHF) range. 
We take advantage of the magnon-photon hybridization between a yttrium iron garnet (YIG) sphere and a copper resonant cavity to employ a resonant heterodyne detection scheme. The measurement is near instantaneous due to the strong coupling attained between magnons and photons. 
In this work measurements are reported showing a significant widening of the measurement bandwidth, obtained by tuning the YIG Larmor frequency with a bias magnetic field and adjusting the magnon-photon coupling strength.
Minimum sensitivity in the sub pT regime is demonstrated in the range 150 -- 225 MHz at room temperature and expected to go to fT in cryogenic temperatures. 
Dynamic range is estimated to be above 100 dB. 
The sensitivity is found to be independent on size, being ready to in-chip miniaturization. 
Such device can be an important building block to quantum circuits, such as baluns, transducers or signal processing units.

\end{abstract}


\maketitle

\section{Introduction}


Ultra low magnetic field detection in the megahertz range is a main component in the study of matter, as in  nuclear magnetic resonance\cite{Reif2021} experiments, and even to axion dark matter searches \cite{Barbieri2017}\cite{Giannotti2025}.
The most straightforward sensors for this applications are induction pick-up coils, since they are wideband up to their resonance frequency, and the induced voltage signal is directly proportional to the measured frequency.
This inherently high sensitity comes with a cost: appropriate electromagnetic shielding and mixing techniques are mandatory to reduce the measurement enviromental noise.
Moreover, the resonance frequency of the circuit is also typically close to very high frequency magnetic fields, degrading signal linearity and sensitivity. 
Thus engineering such pick-up coils can quickly become a very complex problem, decreasing the range of situations where it can be applied\cite{fermon2023}\cite{ripka2021}\cite{asaf2017}.


Some current  alternative ways to measure extremely low ac magnetic fields in the microwave range are: microwave single photon counters\cite{Balembois2024}\cite{Hadfield2009}, optomechanical based magnetometers\cite{Forstner2014}\cite{Forstner2012}\cite{Colombano2020}, and atomic resonance magnetometers\cite{bloom1962Principles}\cite{dupont1969detection}\cite{Budker2007}. 
Although being able to attain incredibly high sensitivities, down to sub fT, they often depend on complex and expensive fabrication procedures and are notoriously narrow banded and with limited dynamic range.

Cavity magnon polaritons (CMP) are the quasi particle produced when the hybridization of magnons and cavity microwave photons are so efficient that perturbation theory breaks down, attaining the polariton state where matter and radiation are indistiguable \cite{ZareRameshti2022}\cite{Purcell1946}.
This state can be attained by using a ferromagnetic medium that overlaps the filling factor of the resonant cavity mode, with both oscillators having sufficiently high quality factors and matched resonance frequencies. 
Thus, in this strong coupling regime, the two oscillators resonances are indistinguable, being phase coherent and sharing the available energy.

In a previous publication\cite{crescini2021}, we presented an ac magnetic probe taking advantage of the CMPs frequency selectivity and phase coherence.
The reported set-up is based on the CMP from the hybridisation occurring between a resonant mode of a microwave cavity and the Kittel mode of a Yttrium Iron Garnet (YIG) sphere placed in a bias magnetic field. 
We were able to achieve a sensitivity of 2 pT/$\sqrt{\mathrm{Hz}}$ over a 1 mm radius cross section, that does not depend on the volume of magnetic material with great potential to in chip miniaturization.
The set-up, however, also suffered from the narrow frequency bandwith issue of the other induction sensors alternative approaches mentioned above.

We present here a follow up of such cavity magnon polariton based field probe, where the measurement bandwidth was widened of about 80 MHz and sensitivity improved to 0.28 pT/$\sqrt{\mathrm{Hz}}$. We also present more detailed study of the sensitivity and dynamic range that can be realized.

\section{Detection setup}

\begin{figure*}[h!]
    \centering
    \includegraphics[width=.85\textwidth]{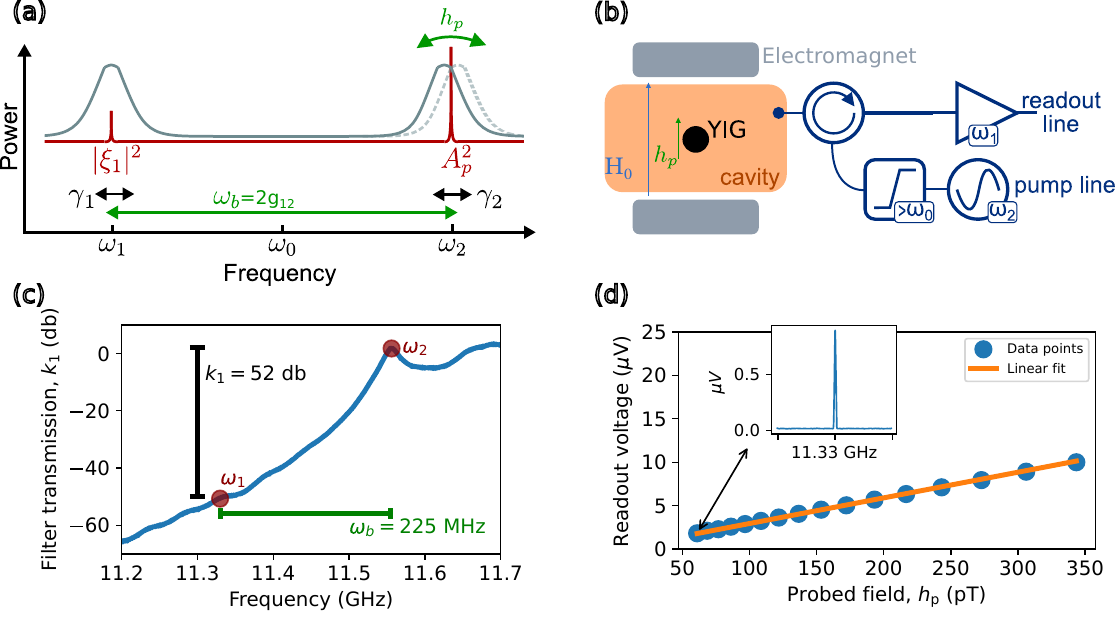}
    \caption{VHF field band probe. a) Hybridization modes $\omega_{1,2}$ with linewidths $\gamma_{1,2}$ separated by $\omega_b$. Adding an ac field at $\omega_b$ generates a sideband $\xi_1$ at $\omega_1$. b) Experimental scheme: the probed field is parallel to the bias field. The microwave pump is filtered to remove its noise. Readout of the power at $\omega_1$ is done by a spectrum analyser after amplification. c) Transmission plot of the waveguide filter showing the 52 dB attenuation at $\omega_1$. d) Sensitivity measurement showing the voltage read at the  spectrum analyser at $\omega_1$ for increasing values of signal fields. A linear fit provides the transduction coefficient of $17.1\pm 2I$ V/mT, and sensitivity is obtained from  the of noise level measurement of 20 nV}
    \label{fig:ExpScheme}
\end{figure*}

A detailed description of the working principle can be found in \cite{crescini2021}, here we will present a schematic version.

The instrument takes advantage of the two modes resulting by the hybridisation between a rectangular microwave cavity mode and the ferromagnetic resonance of a sphere in the strong coupling regime.
The magnetic sphere is placed at the center of a resonant cavity  under a static magnetic field $H_0$. The magnetic field tunes the Larmor resonance of the magnetic material at $\omega_m = \gamma_e H_0$, with $\gamma_e$ the electron gyromagnetic ratio.
In the case of a sphere, excitation at the Larmor frequency corresponds to the fundamental magnetostatic mode known as Kittel mode $\omega_m$, for which uniform precession of the spins occurs. The  coupling $g_{12}$ of the Kittel mode to the rf field of cavity mode at frequency $\omega_0$ is  \cite{wang2018}:

\begin{equation}
    g_{12}(\vec{r})=\frac{\eta_g(\vec{r})}{2}\gamma_e \sqrt{\dfrac{\hbar \omega_0 \mu_0}{V_a}} \sqrt{2N_s},  \label{g12}
\end{equation}
where $\hbar$ is Planck's constant, $\mu_0$ the magnetic permeability of vacuum,   $V_a$ is the resonant mode volume in the cavity and $N_s$ the  number if spins. The geometrical coefficient $\eta_g(\vec{r})$ for a small sphere at the position $\vec{r}$ is:
\begin{equation}
\eta_g (\vec{r}) \simeq \frac{\left[\vec{h}(\vec{r})\cdot\vec{\epsilon}_x\right]^2
+\left[\Vec{h}(\Vec{r})\cdot\Vec{\epsilon}_y\right]^2}{\max \left[\Vec{h} (\Vec{r})\right]^2},  \label{etag}
\end{equation}
where $\vec{h}(\Vec{r})$ is the magnetic component of the cavity rf mode (microwave excitation field) and $\vec{\epsilon}_{x,y}$ are the unitary vectors perpendicular to the static applied field. The value of $\eta_g \leq 1$  represents the  overlap and polarization conditions of the microwave excitation field and the YIG uniform resonance. 

Let's indicate as $\gamma_m$ and $\gamma_0$ the line widths of the Kittel mode and of the cavity mode, respectively. If $(\gamma_m,\gamma_0) < g_{12}$, the system will attain a hybridised state with two new 'hybrid modes' at the resonances $\omega_1<\omega_2$ available:
\begin{equation}
\omega_{1,2}=\frac{\omega_0+\omega_m}{2}\pm \sqrt{\left(\frac{\omega_0-\omega_m}{2}\right)^2+g_{12}^2(\vec{r})}
\label{eq:fdiff}
\end{equation}

It is clear that it is possible to tune the field $H_0$ in such a way that $\omega_m=\omega_0$, condition for the full hybridisation to occur and resulting in $\omega_2 - \omega_1=2g_{12}$.

In the presence of a small ac signal field $h_p(\omega_b)$ at the frequency $\omega_b$, parallel to the bias field $H_0$, phase modulation occurs for the energy stored in the hybrid modes, since following equation (\ref{eq:fdiff}) the signal field will modulate the frequency separation between the two hybrid modes. Specifically, by pumping with a strong pump tone of amplitude  $A_p$ at $\omega_2$, sidebands will be generated at the frequencies $(\omega_2\pm\omega_b)$. If $\omega_b = \omega_2 - \omega_1 = 2 g_{12}$ the lower sideband will be resonant with the lower frequency hybrid mode and can be detected. Its amplitude is proportional to the signal field $h_p$.

The amplitude at the excited  mode resonance frequency $\omega_1$ is then measured via reflection in the resonant cavity, This amplitude is proportional and linear with the amplitude of the magnetic field $h_p$ acting on the magnetic sphere.

The collected power of the signal sideband $P_s$ at $\omega_1$ is given by \cite{crescini2021}:
\begin{align}
P_{s}=\dfrac{\beta_1}{1+\beta_1} \left | \dfrac{\pi \alpha A_p^2 Q_2 h_p}{2 H_0}  \right | +P_{\rm noise}, 
\label{eq:power2b}
\end{align}
where $\beta_1$ is the cavity coupling  at $\omega_1$, $A_p^2$ is the absorbed pump power at $\omega_2$, $Q_2$ the quality factor of $\omega_2$, and $\alpha$ the the field frequency transduction coeficient correction in the gyromagnetic ratio. In the hybridized region, $\omega_{1,2}= \pm \alpha\gamma H_0$, and $\alpha=1/2$ in the fully hybdrized state. The noise power $P_{\rm noise}$ has two main contributors: the thermal noise, and the residual amplitude noise of the pump  ($N_{\rm source}$) at the frequency $\omega_b$ from the carrier. 
The ac magnetic field $h_p$ at $\omega_b$ is then obtained with a measurement of an rf power in a sort of heterodyning scheme. It has to be noted that
Eq. (\ref{eq:power2b}) has no spin quantity parameters: the setup thus can be greatly miniaturised, specially with the help of in chip resonators \cite{Baity2021}.
Our experimental setup (Fig \ref{fig:ExpScheme}b) consists in an electromagnet to provide the static magnetic field and a copper resonant cavity hosting a 2 mm YIG sphere.  A signal generator  provides the pump tone at $\omega_2$, which is filtered with a high-pass filter and fed into the cavity through a circulator. The circulator drives the power reflected from the cavity to a low noise amplifier  (LNA), whose output is read by a spectrum analyser as a voltage over a 50 ohm load.

The copper resonant cavity is $32$ mm long and has quasi-rectangular cross section of $14.8 \times 24.8$ mm$^2$. The selected cavity mode is a TE102/TM102 one \footnote{In our setup geometry, the magnetic field in the center of the cavity for both modes has the adequate perpendicular directions considering the volume of the sample.}, where the cavity resonance frequency is $\omega_0 /2 \pi=11.43$ GHz with a quality factor of above 5000. At bias field $H_0=\omega_0/ \gamma_e \approx 0.4$ T the system is hybridized with $\omega_1/2\pi= 11.33$ GHz  and $\omega_2 / 2 \pi=11.55$ GHz, with $\gamma_1\approx\gamma_2=1.5$ MHz. The resulting coupling is of $2g_{12}=\omega_2-\omega_1 \approx (2\pi\, 220$ MHz) close to full hybridisation, i.e. with $\alpha \approx 0.5$.

The pump tone $A_p$ is provided by a signal generator having maximum output of 16 dBm. In order to remove as much as possible its residual amplitude modulation (RAM) at the frequency $\omega_1$, the pump is filtered by a waveguide. With a cross  section of $12.8  \times 6.3 $ mm$^2$, the waveguide has an attenuation of $k_1\simeq50$ dB at  $\omega_1$ with respect to $\omega_2$. Thanks to the filtering the only contribution to $P_{\rm noise}$ is the thermal one.

Test fields for the magnetometer can be fed by means of a 6 mm diameter coil (SC - Sensing Coil) mounted in a 1 cm diameter hole placed on a side wall of the cavity. This small coil is producing almost uniform magnetic field on the YIG sample, with a direction parallel to the external bias field $H_0$. Test field can be generated driving the coil with a function generator or connecting the coil to an external loop for field pick up. The magnetic field generated by the sensing coil can be calibrated by sending a dc current on the coil and measuring the change of the Larmor frequency in the YIG sphere.


\subsection{Sensitivity}

The sensitivity of the setup, defined as the minimum signal measurable at  signal to noise ratio of one is given by \cite{crescini2021}:

\begin{align}
\sigma_{h_{p}}&=\frac{2\eta H_{0}}{\pi\alpha Q_{2}^{l}}\left(\sqrt{\frac{k_{B}T_{s}}{A_{P}^{2}}+N_{\rm source}}\right),\label{eq:sens}
\end{align} 

where:
\begin{align}
N_{\rm source}&=k_{1}\left(\dfrac{1-\beta_{1}}{1+\beta_{1}}\right)^{2}|{\rm RAM}(\omega_{b})|^{2}  \label{ram} \\
\eta&= \dfrac{1+\beta_2}{2}\sqrt{\dfrac{1+\beta_1}{\beta_2\beta_1}}\\
T_s &= \dfrac{4\beta_1 T_c}{(1+ \beta_1^2)} + T_n\label{eq:sensdet}
\end{align} $\beta_{1,2}$ are the antenna coupling coefficients of the hybridized modes, $k_B$ the Boltzmann constant, , $Q^l_2$ the loaded quality factor of the mode at $\omega_2$, $A_p^2$ the microwave pump power, $k_1$ the transmission coefficient of the filter, $T_c$ the thermodynamic temperature of the system, $T_n$ the noise temperature of the low noise amplifier.

The ultimate noise level is essentially given by the thermal fluctuations and amplifier added noise, at least when operating at room temperature. In fact, for our signal generator it is found that the residual amplitude modulation at the signal frequency is $|{\rm RAM}(\omega_b)|^2\approx\,160 $ dBc/Hz. Giving the attenuation of the filter, this is well below the thermal term.

Experimentally, the sensitivity is obtained in a two step process. First, the transduction coefficient from test field to readout voltage is measured by sending different values of test field on the YIG sphere. With a signal generator it is indeed possible to drive the single loop coil SC and generate a known magnetic field on the YIG sphere as explained in the previous section. In figure \ref{fig:ExpScheme}(d) a graph shows the voltage measured on the spectrum analyser varying the field $h_p$, the slope of the linear fit is the transduction coefficient. In the second step a long measurement is performed without any input on the coil SC in order to obtain a good estimate of the background noise. Sensitivity is then obtained by dividing the background noise by the transduction coefficient. More details on the procedure can be found in \cite{crescini2021}.


\section{Improved setup}

\subsection{Improving bandwidth}

With the sphere placed at the cavity center, where a maximum of the  rf magnetic field is present, the  magnetometer can only sense ac fields at the frequency corresponding to $2 g_{12}$. We can use Equation (\ref{eq:fdiff}) to write in a more useful way the working frequency $\omega_s$  of the apparatus:

\begin{equation}
\omega_s = \sqrt{\left[ \frac{\omega_0 - \omega_m(H_0)}{2}\right]^2 + \left[g_{12,\rm fh}(\vec{r})\right]^2}
\end{equation}

where we have defined $g_{12,\rm fh}(\vec{r})=g_{12}(\vec{r})/\eta_g(\vec{r})$, i.e. the coupling in the presence of full hybridisation for which $\eta_g=1$.

It is possible to change $\omega_s$ in two ways:

\begin{description}
\item[a] increase $\omega_s$ by changing the bias field $H_0$ \cite{crescini2020magnon}. This will produce an asymmetric hybridisation and the coupling of the two modes will be different.
\item[b] decrease $\omega_s$ by changing the position of the sphere from cavity center. In this case the system will always work in the regime of full hybridisation with symmetric modes.
\end{description}

We thus use a combination of the two strategies to probe different value of $\omega_b$: slightly increasing/decreasing the bias magnetic field out of the full hybridization field and moving the YIG sphere out of the center of the cavity (where the maximum of the TE102 mode is located) of a few mm. When performing these operations, however, one has to consider the transmission function for the microwave components of the system, the most critical one being the waveguide used for filtering. The best way to handle this is to measure the reflection function of the system, i.e. going from the pump line to the readout line through the circulator (see Figure \ref{fig:ExpScheme}(b)). Such a measurement will show the mixed effect of the changing hybrid mode frequencies and waveguide filter.

\begin{figure}[h!]
    \centering
    \includegraphics[width=1\linewidth]{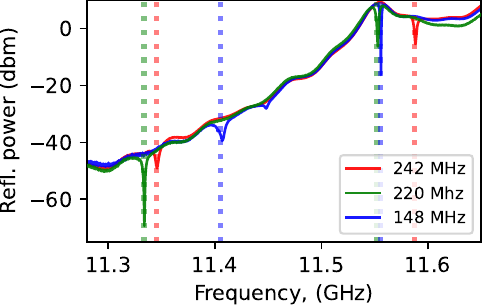}
    \caption{Effect of changing the coupling constants in the hybridization modes. Decreasing the coupling constant also reduces the effectivity of the high pass filter and $A_p^2$. The slope on the left is the roll-off of the waveguide high-pass filter.}
    \label{fig:bandReflec}
\end{figure}

In figure \ref{fig:bandReflec} one can see the reflection functions for three representative choices of frequencies $\omega_s$. The green curve is the 'standard' one obtained with the sphere at the cavity center, $\omega_c = \omega_m$, i.e. $\omega_s/2\pi=220$ MHz. To increase from the standard value, the only option is to change the static applied field $H_0$: in order to keep a good transmissivity at the frequency $\omega_2$, with a necessary strong pump, $H_0$ is increased but the filtering is as well less efficient since we are going above a  peak ripple oscillations from the filter itself (Red curve in figure with $\omega_s/2\pi=242 $ MHz).  To sense lower values for $\omega_s$, a combination of position of the sphere and raising of the bias magnetic field is chosen in order to enjoy the best position in the filter transfer function (Blue curve in figure with $\omega_s/2\pi=148 $ MHz).

\begin{figure}[h!]
    \centering
    \includegraphics[width=1\linewidth]{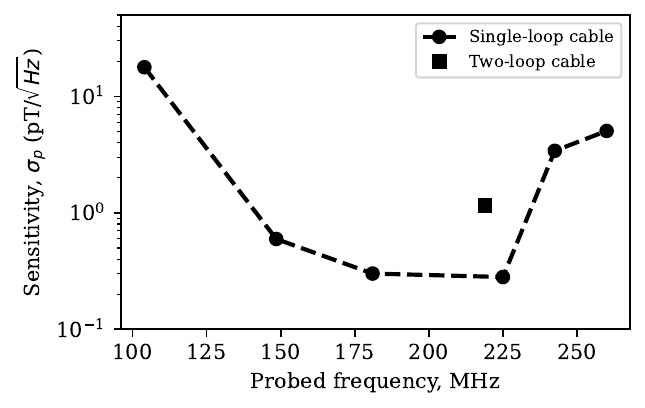}
    \caption{Summary of the sensitivity measurements for different probing frequencies (coupling constants). The square symbol is the extended cable measurements.}
    \label{fig:bandSumm}
\end{figure}

In figure \ref{fig:bandSumm} is shown the measured sensitivities of diverse probed frequencies $\omega_b$ for $h_p$. The pump power was kept constant at $A_p^2=5$ mW (7dBm) and $\beta_{1,2}$ as close as possible to $1$ during all measurements.   The background noise increases substantially above 220 MHz, while the sensitivity has a plateau with optimal operation between 148 and 225 MHz where sub pT sensitivity is reached.   

The diverse positions of $\omega_{1,2}$ around the filter roll-off reduces its effectiveness in removing $N_{\rm source}$ and providing the pump power $A_p^2$ inside the cavity. In the 115 MHz case, the filter is still able to reduce $N_{\rm source}$ to negligible levels, however $A_p^2$ is also heavily filtered. Above 220 MHz, the $A_p^2$ is reduced due to the ripple fluctuations, and $N_{\rm source}$ starts to pass through and increase the background noise. 

Figure \ref{fig:bandSumm} also shows a measurement performed by using a two-loop cable to pick up the signal from the ambient. To decrease impedance mismatch, the cable has two identical single turn 6 mm diameter coils at each end, and it is 1 m long. One end is placed in the cavity side hole to act on the YIG sphere, the other end probes the field generated by a coil connected to the function generator. We have verified that there were less than $O(1)$ sensitivity loss using this 1 m long braided cable between the two loops. The sensitivity at optimum conditions decreased from $\sigma_{b1}=281$ fT$/\sqrt{\rm Hz}$ to  $\sigma_{b1}=1.16$ pT$/\sqrt{\rm Hz}$, without accounting for cable losses. The braided cable provides a simple way to introduce far away signals inside the cavity, but is also serving as the first stage of detection. We could not measure the signal provided by the braided cable at our oscilloscope with similar coils. The setup in this case acts more as an amplifier and signal processing unit.



\subsection{Improving sensitivity and dynamic range}

In order to maximise the sensitivity, a compromise between the parameters of equation (\ref{eq:sens}) has to be found. 
\begin{itemize}
\item The coupling coefficients $\beta_{1,2}$ have a value close to 1 in our setup and  the coefficient $\eta$ acts for a factor of less than three in the best possible $\beta_{1,2}$ combination. 
\item As long as the filter used is steep enough, $N_{\rm source}$ can be neglected. 
\item The hybridization slope constant $\alpha$ changes with the field, but has a maximum of 1/2 and changes the amplitude and $Q$ of the modes.  
\item As long as no cryogenic solutions are employed, the noise temperature $T_s$ of the system can be improved only by the use of a good quality low noise amplifier. 
\item The quality factor of the modes are limited by the resonant cavity quality factor  and the YIG Gilbert damping constant. The first can be improved in standard oxygen free copper cavities up to $10^5$\cite{spencer1959}, and higher values can be attained with  superconductor on chip resonators\cite{Baity2021}.Nevertheless the hybridized quality factor is ultimately limited by the magnetic material, and the state of the art, with YIG spheres,  is in the order of  $10^{5}$\cite{spencer1959}.  
\item The major parameter to optimize thus remains the absorbed power, and its relationship with the aforementioned parameters.
\end{itemize}

Increasing the power in the cavity raises the exciting rf field, and eventually spin wave instabilities in the YIG sphere will be present above non-linear threshold fields. These effects have been studied extensively by Suhl and others \cite{SUHL1957209,bryant1988,rezende1990}. More specifically, at higher resonance fields, second order relaxation process are expected, limiting the amplitude growth of the main resonance peak and enlarging the linewidth. However, in the strong magnon-photon coupling regime, the cavity offers extra dissipation channels, and only a few studies exists  in the literature. Lee et al\cite{kurebayashi2023} managed to break the hybridization of a bulk YIG sample and a loop resonator by increasing the excitation power; Wang et al \cite{wang2018} found instabilities above 20 dBm excitation power for a 1 mm diameter YIG sphere in a $Q\approx 2000$ cavity, resulting in a frequency shift of tens of MHz due to magnetic shape anisotropy.

Nonetheless, field and frequency modulation techniques are known to increase the instability thresholds, analogous to beam stabilization in the time domain\cite{hartwig_suppression_1967}. Due to the long magnon growth time of parametric processes, these signal modulations changes the dispersion relation of the spin waves before they can populate, increasing the threshold fields. It has been shown that field modulation on the range of kHz allows for great magnetic precession amplitudes without generating instabilities. 

In the present setup, the probed field already acts as a field modulation. Considering the susceptibility discontinuity condition of a macrospin model, $h_x \leq H_{0}/2Q^{l}_2 $, where $h_x$ is the excitation field provided by the cavity, one can estimate the maximum pumping power possible  in Equation (\ref{eq:sens}).
In other words, the maximum power $A_p^2$ in Eq. \ref{eq:sens} is limited by the maximum precession angle that can be achieved by the magnetization dynamics of the YIG sphere.
For $ Q^{l}_2=9000$, the limit is found at 40 dBm or 10 W.  
The ultimate lower bound for sensitivity can then be estimated. For different operating temperatures, at $0.4$ T, $\sigma_{h_p}$ decreases from $107$ fT at $300$ K (room temperature) to $66$ fT at $87$ K (nitrogen) and finally $16$ fT at $4$ K (helium). Our best sensitivity measured  to date at room temperature and similar parameters is $281$ fT/$\sqrt{Hz}$ with $A_p^2=5$ mW, improving our old value of 1.9 pT/$\sqrt{Hz}$ \cite{crescini2021}.

The dynamic range of the setup is a function of both sensitivity and the maximum allowed $h_p$. For the later, due to to the orientation of the fields, parallel pumping is considered, exciting magnons at half the frequency of the probing field acting as pump. The working frequencies are of hundreds of MHz, far from the 5 GHz needed at 400 mT. Moreover, for YIG spheres, the parallel pumping thresholds are high due to the spherical symmetry and lack of shape anisotropy. Thus, the maximum probing field $h_p$ allowed has to be comparable to the bias field to actually cause instabilities. Taking a conservative $h_p=1$ mT, it is expected a 100 dB of dynamic range at room temperature.

\section{Conclusions}
We demonstrated that our polariton based magnetic field probe is able to maintain its sub pico Tesla sensitivity in a range of a little less than a hundred MHz with more than 100 Db dynamic range. The field is probed only around the diameter of the YIG sphere, localized at 2mm diameter radius. 

The setup can be customized for a variety of applications. The operation frequency can be tuned with the coupling, as a function of magnetic volume and cavity filling factor. The bandwidth can be chosen with small variations of the applied field and inhomogeneity of the excitation field provided by the cavity. The probed field can be measured directly via a hole in the cavity or by introducing a simple impedance matched cable. Due to the independence of sensitivity with magnetic volume, miniaturization is possible employing planar resonators\cite{kobe_review_2017}\cite{Baity2021} and micro-electromechanical systems\cite{cocconcelli2024}.

The efficient field to frequency convertion due to the sidebands produced around the magnon polariton modes can also be exploited for developing supporting tuneable devices for quantum electronics circuits, such as transducers and amplifiers.

\section*{Acknowledgments}

We are grateful to E. Berto for  mechanical work; F. Calaon and M. Tessaro for help with the electronics. We are also greateful to C. Fermon and G. De Loubens for the critical insights. 

\bibliography{BSPaper.bib}

\begin{thebibliography}{29}%
\makeatletter
\providecommand \@ifxundefined [1]{%
 \@ifx{#1\undefined}
}%
\providecommand \@ifnum [1]{%
 \ifnum #1\expandafter \@firstoftwo
 \else \expandafter \@secondoftwo
 \fi
}%
\providecommand \@ifx [1]{%
 \ifx #1\expandafter \@firstoftwo
 \else \expandafter \@secondoftwo
 \fi
}%
\providecommand \natexlab [1]{#1}%
\providecommand \enquote  [1]{``#1''}%
\providecommand \bibnamefont  [1]{#1}%
\providecommand \bibfnamefont [1]{#1}%
\providecommand \citenamefont [1]{#1}%
\providecommand \href@noop [0]{\@secondoftwo}%
\providecommand \href [0]{\begingroup \@sanitize@url \@href}%
\providecommand \@href[1]{\@@startlink{#1}\@@href}%
\providecommand \@@href[1]{\endgroup#1\@@endlink}%
\providecommand \@sanitize@url [0]{\catcode `\\12\catcode `\$12\catcode
  `\&12\catcode `\#12\catcode `\^12\catcode `\_12\catcode `\%12\relax}%
\providecommand \@@startlink[1]{}%
\providecommand \@@endlink[0]{}%
\providecommand \url  [0]{\begingroup\@sanitize@url \@url }%
\providecommand \@url [1]{\endgroup\@href {#1}{\urlprefix }}%
\providecommand \urlprefix  [0]{URL }%
\providecommand \Eprint [0]{\href }%
\providecommand \doibase [0]{https://doi.org/}%
\providecommand \selectlanguage [0]{\@gobble}%
\providecommand \bibinfo  [0]{\@secondoftwo}%
\providecommand \bibfield  [0]{\@secondoftwo}%
\providecommand \translation [1]{[#1]}%
\providecommand \BibitemOpen [0]{}%
\providecommand \bibitemStop [0]{}%
\providecommand \bibitemNoStop [0]{.\EOS\space}%
\providecommand \EOS [0]{\spacefactor3000\relax}%
\providecommand \BibitemShut  [1]{\csname bibitem#1\endcsname}%
\let\auto@bib@innerbib\@empty
\bibitem [{\citenamefont {Reif}\ \emph {et~al.}(2021)\citenamefont {Reif},
  \citenamefont {Ashbrook}, \citenamefont {Emsley},\ and\ \citenamefont
  {Hong}}]{Reif2021}%
  \BibitemOpen
  \bibfield  {author} {\bibinfo {author} {\bibfnamefont {B.}~\bibnamefont
  {Reif}}, \bibinfo {author} {\bibfnamefont {S.~E.}\ \bibnamefont {Ashbrook}},
  \bibinfo {author} {\bibfnamefont {L.}~\bibnamefont {Emsley}},\ and\ \bibinfo
  {author} {\bibfnamefont {M.}~\bibnamefont {Hong}},\ }\bibfield  {journal}
  {\bibinfo  {journal} {Nature Reviews Methods Primers}\ }\textbf {\bibinfo
  {volume} {1}},\ \href
  {https://doi.org/https://doi.org/10.1038/s43586-020-00002-1}
  {https://doi.org/10.1038/s43586-020-00002-1} (\bibinfo {year}
  {2021})\BibitemShut {NoStop}%
\bibitem [{\citenamefont {Barbieri}\ \emph {et~al.}(2017)\citenamefont
  {Barbieri}, \citenamefont {Braggio}, \citenamefont {Carugno}, \citenamefont
  {Gallo}, \citenamefont {Lombardi}, \citenamefont {Ortolan}, \citenamefont
  {Pengo}, \citenamefont {Ruoso},\ and\ \citenamefont {Speake}}]{Barbieri2017}%
  \BibitemOpen
  \bibfield  {author} {\bibinfo {author} {\bibfnamefont {R.}~\bibnamefont
  {Barbieri}}, \bibinfo {author} {\bibfnamefont {C.}~\bibnamefont {Braggio}},
  \bibinfo {author} {\bibfnamefont {G.}~\bibnamefont {Carugno}}, \bibinfo
  {author} {\bibfnamefont {C.}~\bibnamefont {Gallo}}, \bibinfo {author}
  {\bibfnamefont {A.}~\bibnamefont {Lombardi}}, \bibinfo {author}
  {\bibfnamefont {A.}~\bibnamefont {Ortolan}}, \bibinfo {author} {\bibfnamefont
  {R.}~\bibnamefont {Pengo}}, \bibinfo {author} {\bibfnamefont
  {G.}~\bibnamefont {Ruoso}},\ and\ \bibinfo {author} {\bibfnamefont
  {C.}~\bibnamefont {Speake}},\ }\href
  {https://doi.org/10.1016/j.dark.2017.01.003} {\bibfield  {journal} {\bibinfo
  {journal} {Physics of the Dark Universe}\ }\textbf {\bibinfo {volume} {15}},\
  \bibinfo {pages} {135} (\bibinfo {year} {2017})}\BibitemShut {NoStop}%
\bibitem [{\citenamefont {Giannotti}(2025)}]{Giannotti2025}%
  \BibitemOpen
  \bibfield  {author} {\bibinfo {author} {\bibfnamefont {M.}~\bibnamefont
  {Giannotti}},\ }in\ \href {https://doi.org/10.22323/1.474.0033} {\emph
  {\bibinfo {booktitle} {Proceedings of 2nd Training School and General Meeting
  of the COST Action COSMIC WISPers (CA21106) — PoS(COSMICWISPers2024)}}},\
  \bibinfo {series and number} {COSMICWISPers2024}\ (\bibinfo  {publisher}
  {Sissa Medialab},\ \bibinfo {year} {2025})\ p.\ \bibinfo {pages}
  {033}\BibitemShut {NoStop}%
\bibitem [{\citenamefont {Fermon}(2017)}]{fermon2023}%
  \BibitemOpen
  \bibfield  {author} {\bibinfo {author} {\bibfnamefont {C.}~\bibnamefont
  {Fermon}},\ }\bibinfo {title} {Introduction on magnetic sensing and spin
  electronics},\ in\ \href
  {https://doi.org/https://doi.org/10.1002/9783527698509.ch1} {\emph {\bibinfo
  {booktitle} {Nanomagnetism: Applications and Perspectives}}}\ (\bibinfo
  {publisher} {John Wiley Sons, Ltd},\ \bibinfo {year} {2017})\ Chap.~\bibinfo
  {chapter} {1}, pp.\ \bibinfo {pages} {1--18}\BibitemShut {NoStop}%
\bibitem [{\citenamefont {Ripka}(2021)}]{ripka2021}%
  \BibitemOpen
  \bibfield  {author} {\bibinfo {author} {\bibfnamefont {P.}~\bibnamefont
  {Ripka}},\ }\href@noop {} {\emph {\bibinfo {title} {Magnetic Sensors and
  Magnetometers, Second Edition}}}\ (\bibinfo  {publisher} {Artech},\ \bibinfo
  {year} {2021})\BibitemShut {NoStop}%
\bibitem [{\citenamefont {Asaf~Grosz}(2017)}]{asaf2017}%
  \BibitemOpen
  \bibfield  {author} {\bibinfo {author} {\bibfnamefont {S.~C.~M.}\
  \bibnamefont {Asaf~Grosz}, \bibfnamefont {Michael J. Haji-Sheikh}},\ }\href
  {https://doi.org/10.1007/978-3-319-34070-8} {\emph {\bibinfo {title} {High
  Sensitivity Magnetometers}}}\ (\bibinfo  {publisher} {Springer International
  Publishing},\ \bibinfo {year} {2017})\BibitemShut {NoStop}%
\bibitem [{\citenamefont {Balembois}\ \emph {et~al.}(2024)\citenamefont
  {Balembois}, \citenamefont {Travesedo}, \citenamefont {Pallegoix},
  \citenamefont {May}, \citenamefont {Billaud}, \citenamefont {Villiers},
  \citenamefont {Estève}, \citenamefont {Vion}, \citenamefont {Bertet},\ and\
  \citenamefont {Flurin}}]{Balembois2024}%
  \BibitemOpen
  \bibfield  {author} {\bibinfo {author} {\bibfnamefont {L.}~\bibnamefont
  {Balembois}}, \bibinfo {author} {\bibfnamefont {J.}~\bibnamefont
  {Travesedo}}, \bibinfo {author} {\bibfnamefont {L.}~\bibnamefont
  {Pallegoix}}, \bibinfo {author} {\bibfnamefont {A.}~\bibnamefont {May}},
  \bibinfo {author} {\bibfnamefont {E.}~\bibnamefont {Billaud}}, \bibinfo
  {author} {\bibfnamefont {M.}~\bibnamefont {Villiers}}, \bibinfo {author}
  {\bibfnamefont {D.}~\bibnamefont {Estève}}, \bibinfo {author} {\bibfnamefont
  {D.}~\bibnamefont {Vion}}, \bibinfo {author} {\bibfnamefont {P.}~\bibnamefont
  {Bertet}},\ and\ \bibinfo {author} {\bibfnamefont {E.}~\bibnamefont
  {Flurin}},\ }\href {https://doi.org/10.1103/physrevapplied.21.014043}
  {\bibfield  {journal} {\bibinfo  {journal} {Physical Review Applied}\
  }\textbf {\bibinfo {volume} {21}},\ \bibinfo {pages} {014043} (\bibinfo
  {year} {2024})}\BibitemShut {NoStop}%
\bibitem [{\citenamefont {Hadfield}(2009)}]{Hadfield2009}%
  \BibitemOpen
  \bibfield  {author} {\bibinfo {author} {\bibfnamefont {R.~H.}\ \bibnamefont
  {Hadfield}},\ }\href {https://doi.org/10.1038/nphoton.2009.230} {\bibfield
  {journal} {\bibinfo  {journal} {Nature Photonics}\ }\textbf {\bibinfo
  {volume} {3}},\ \bibinfo {pages} {696} (\bibinfo {year} {2009})}\BibitemShut
  {NoStop}%
\bibitem [{\citenamefont {Forstner}\ \emph {et~al.}(2014)\citenamefont
  {Forstner}, \citenamefont {Sheridan}, \citenamefont {Knittel}, \citenamefont
  {Humphreys}, \citenamefont {Brawley}, \citenamefont {Rubinsztein‐Dunlop},\
  and\ \citenamefont {Bowen}}]{Forstner2014}%
  \BibitemOpen
  \bibfield  {author} {\bibinfo {author} {\bibfnamefont {S.}~\bibnamefont
  {Forstner}}, \bibinfo {author} {\bibfnamefont {E.}~\bibnamefont {Sheridan}},
  \bibinfo {author} {\bibfnamefont {J.}~\bibnamefont {Knittel}}, \bibinfo
  {author} {\bibfnamefont {C.~L.}\ \bibnamefont {Humphreys}}, \bibinfo {author}
  {\bibfnamefont {G.~A.}\ \bibnamefont {Brawley}}, \bibinfo {author}
  {\bibfnamefont {H.}~\bibnamefont {Rubinsztein‐Dunlop}},\ and\ \bibinfo
  {author} {\bibfnamefont {W.~P.}\ \bibnamefont {Bowen}},\ }\href
  {https://doi.org/10.1002/adma.201401144} {\bibfield  {journal} {\bibinfo
  {journal} {Advanced Materials}\ }\textbf {\bibinfo {volume} {26}},\ \bibinfo
  {pages} {6348} (\bibinfo {year} {2014})}\BibitemShut {NoStop}%
\bibitem [{\citenamefont {Forstner}\ \emph {et~al.}(2012)\citenamefont
  {Forstner}, \citenamefont {Prams}, \citenamefont {Knittel}, \citenamefont
  {van Ooijen}, \citenamefont {Swaim}, \citenamefont {Harris}, \citenamefont
  {Szorkovszky}, \citenamefont {Bowen},\ and\ \citenamefont
  {Rubinsztein-Dunlop}}]{Forstner2012}%
  \BibitemOpen
  \bibfield  {author} {\bibinfo {author} {\bibfnamefont {S.}~\bibnamefont
  {Forstner}}, \bibinfo {author} {\bibfnamefont {S.}~\bibnamefont {Prams}},
  \bibinfo {author} {\bibfnamefont {J.}~\bibnamefont {Knittel}}, \bibinfo
  {author} {\bibfnamefont {E.~D.}\ \bibnamefont {van Ooijen}}, \bibinfo
  {author} {\bibfnamefont {J.~D.}\ \bibnamefont {Swaim}}, \bibinfo {author}
  {\bibfnamefont {G.~I.}\ \bibnamefont {Harris}}, \bibinfo {author}
  {\bibfnamefont {A.}~\bibnamefont {Szorkovszky}}, \bibinfo {author}
  {\bibfnamefont {W.~P.}\ \bibnamefont {Bowen}},\ and\ \bibinfo {author}
  {\bibfnamefont {H.}~\bibnamefont {Rubinsztein-Dunlop}},\ }\href
  {https://doi.org/10.1103/physrevlett.108.120801} {\bibfield  {journal}
  {\bibinfo  {journal} {Physical Review Letters}\ }\textbf {\bibinfo {volume}
  {108}},\ \bibinfo {pages} {120801} (\bibinfo {year} {2012})}\BibitemShut
  {NoStop}%
\bibitem [{\citenamefont {Colombano}\ \emph {et~al.}(2020)\citenamefont
  {Colombano}, \citenamefont {Arregui}, \citenamefont {Bonell}, \citenamefont
  {Capuj}, \citenamefont {Chavez-Angel}, \citenamefont {Pitanti}, \citenamefont
  {Valenzuela}, \citenamefont {Sotomayor-Torres}, \citenamefont
  {Navarro-Urrios},\ and\ \citenamefont {Costache}}]{Colombano2020}%
  \BibitemOpen
  \bibfield  {author} {\bibinfo {author} {\bibfnamefont {M.}~\bibnamefont
  {Colombano}}, \bibinfo {author} {\bibfnamefont {G.}~\bibnamefont {Arregui}},
  \bibinfo {author} {\bibfnamefont {F.}~\bibnamefont {Bonell}}, \bibinfo
  {author} {\bibfnamefont {N.}~\bibnamefont {Capuj}}, \bibinfo {author}
  {\bibfnamefont {E.}~\bibnamefont {Chavez-Angel}}, \bibinfo {author}
  {\bibfnamefont {A.}~\bibnamefont {Pitanti}}, \bibinfo {author} {\bibfnamefont
  {S.}~\bibnamefont {Valenzuela}}, \bibinfo {author} {\bibfnamefont
  {C.}~\bibnamefont {Sotomayor-Torres}}, \bibinfo {author} {\bibfnamefont
  {D.}~\bibnamefont {Navarro-Urrios}},\ and\ \bibinfo {author} {\bibfnamefont
  {M.}~\bibnamefont {Costache}},\ }\href
  {https://doi.org/10.1103/physrevlett.125.147201} {\bibfield  {journal}
  {\bibinfo  {journal} {Physical Review Letters}\ }\textbf {\bibinfo {volume}
  {125}},\ \bibinfo {pages} {147201} (\bibinfo {year} {2020})}\BibitemShut
  {NoStop}%
\bibitem [{\citenamefont {Bloom}(1962)}]{bloom1962Principles}%
  \BibitemOpen
  \bibfield  {author} {\bibinfo {author} {\bibfnamefont {A.~L.}\ \bibnamefont
  {Bloom}},\ }\href@noop {} {\bibfield  {journal} {\bibinfo  {journal} {Applied
  Optics}\ }\textbf {\bibinfo {volume} {1}},\ \bibinfo {pages} {61} (\bibinfo
  {year} {1962})}\BibitemShut {NoStop}%
\bibitem [{\citenamefont {Dupont-Roc}\ \emph {et~al.}(1969)\citenamefont
  {Dupont-Roc}, \citenamefont {Haroche},\ and\ \citenamefont
  {Cohen-Tannoudji}}]{dupont1969detection}%
  \BibitemOpen
  \bibfield  {author} {\bibinfo {author} {\bibfnamefont {J.}~\bibnamefont
  {Dupont-Roc}}, \bibinfo {author} {\bibfnamefont {S.}~\bibnamefont
  {Haroche}},\ and\ \bibinfo {author} {\bibfnamefont {C.}~\bibnamefont
  {Cohen-Tannoudji}},\ }\href@noop {} {\bibfield  {journal} {\bibinfo
  {journal} {Physics Letters A}\ }\textbf {\bibinfo {volume} {28}},\ \bibinfo
  {pages} {638} (\bibinfo {year} {1969})}\BibitemShut {NoStop}%
\bibitem [{\citenamefont {Budker}\ and\ \citenamefont
  {Romalis}(2007)}]{Budker2007}%
  \BibitemOpen
  \bibfield  {author} {\bibinfo {author} {\bibfnamefont {D.}~\bibnamefont
  {Budker}}\ and\ \bibinfo {author} {\bibfnamefont {M.}~\bibnamefont
  {Romalis}},\ }\href {https://doi.org/10.1038/nphys566} {\bibfield  {journal}
  {\bibinfo  {journal} {Nature Physics}\ }\textbf {\bibinfo {volume} {3}},\
  \bibinfo {pages} {227} (\bibinfo {year} {2007})}\BibitemShut {NoStop}%
\bibitem [{\citenamefont {Zare~Rameshti}\ \emph {et~al.}(2022)\citenamefont
  {Zare~Rameshti}, \citenamefont {Viola~Kusminskiy}, \citenamefont {Haigh},
  \citenamefont {Usami}, \citenamefont {Lachance-Quirion}, \citenamefont
  {Nakamura}, \citenamefont {Hu}, \citenamefont {Tang}, \citenamefont {Bauer},\
  and\ \citenamefont {Blanter}}]{ZareRameshti2022}%
  \BibitemOpen
  \bibfield  {author} {\bibinfo {author} {\bibfnamefont {B.}~\bibnamefont
  {Zare~Rameshti}}, \bibinfo {author} {\bibfnamefont {S.}~\bibnamefont
  {Viola~Kusminskiy}}, \bibinfo {author} {\bibfnamefont {J.~A.}\ \bibnamefont
  {Haigh}}, \bibinfo {author} {\bibfnamefont {K.}~\bibnamefont {Usami}},
  \bibinfo {author} {\bibfnamefont {D.}~\bibnamefont {Lachance-Quirion}},
  \bibinfo {author} {\bibfnamefont {Y.}~\bibnamefont {Nakamura}}, \bibinfo
  {author} {\bibfnamefont {C.-M.}\ \bibnamefont {Hu}}, \bibinfo {author}
  {\bibfnamefont {H.~X.}\ \bibnamefont {Tang}}, \bibinfo {author}
  {\bibfnamefont {G.~E.}\ \bibnamefont {Bauer}},\ and\ \bibinfo {author}
  {\bibfnamefont {Y.~M.}\ \bibnamefont {Blanter}},\ }\href
  {https://doi.org/10.1016/j.physrep.2022.06.001} {\bibfield  {journal}
  {\bibinfo  {journal} {Physics Reports}\ }\textbf {\bibinfo {volume} {979}},\
  \bibinfo {pages} {1} (\bibinfo {year} {2022})}\BibitemShut {NoStop}%
\bibitem [{\citenamefont {Purcell}\ \emph {et~al.}(1946)\citenamefont
  {Purcell}, \citenamefont {Torrey},\ and\ \citenamefont
  {Pound}}]{Purcell1946}%
  \BibitemOpen
  \bibfield  {author} {\bibinfo {author} {\bibfnamefont {E.~M.}\ \bibnamefont
  {Purcell}}, \bibinfo {author} {\bibfnamefont {H.~C.}\ \bibnamefont
  {Torrey}},\ and\ \bibinfo {author} {\bibfnamefont {R.~V.}\ \bibnamefont
  {Pound}},\ }\href {https://doi.org/10.1103/physrev.69.37} {\bibfield
  {journal} {\bibinfo  {journal} {Physical Review}\ }\textbf {\bibinfo {volume}
  {69}},\ \bibinfo {pages} {37} (\bibinfo {year} {1946})}\BibitemShut {NoStop}%
\bibitem [{\citenamefont {Crescini}\ \emph {et~al.}(2021)\citenamefont
  {Crescini}, \citenamefont {Carugno},\ and\ \citenamefont
  {Ruoso}}]{crescini2021}%
  \BibitemOpen
  \bibfield  {author} {\bibinfo {author} {\bibfnamefont {N.}~\bibnamefont
  {Crescini}}, \bibinfo {author} {\bibfnamefont {G.}~\bibnamefont {Carugno}},\
  and\ \bibinfo {author} {\bibfnamefont {G.}~\bibnamefont {Ruoso}},\ }\href
  {https://doi.org/10.1103/PhysRevApplied.16.034036} {\bibfield  {journal}
  {\bibinfo  {journal} {Phys. Rev. Applied}\ }\textbf {\bibinfo {volume}
  {16}},\ \bibinfo {pages} {034036} (\bibinfo {year} {2021})}\BibitemShut
  {NoStop}%
\bibitem [{\citenamefont {Wang}\ \emph {et~al.}(2018)\citenamefont {Wang},
  \citenamefont {Zhang}, \citenamefont {Zhang}, \citenamefont {Li},
  \citenamefont {Hu},\ and\ \citenamefont {You}}]{wang2018}%
  \BibitemOpen
  \bibfield  {author} {\bibinfo {author} {\bibfnamefont {Y.-P.}\ \bibnamefont
  {Wang}}, \bibinfo {author} {\bibfnamefont {G.-Q.}\ \bibnamefont {Zhang}},
  \bibinfo {author} {\bibfnamefont {D.}~\bibnamefont {Zhang}}, \bibinfo
  {author} {\bibfnamefont {T.-F.}\ \bibnamefont {Li}}, \bibinfo {author}
  {\bibfnamefont {C.-M.}\ \bibnamefont {Hu}},\ and\ \bibinfo {author}
  {\bibfnamefont {J.~Q.}\ \bibnamefont {You}},\ }\href
  {https://doi.org/10.1103/PhysRevLett.120.057202} {\bibfield  {journal}
  {\bibinfo  {journal} {Phys. Rev. Lett.}\ }\textbf {\bibinfo {volume} {120}},\
  \bibinfo {pages} {057202} (\bibinfo {year} {2018})}\BibitemShut {NoStop}%
\bibitem [{\citenamefont {Baity}\ \emph {et~al.}(2021)\citenamefont {Baity},
  \citenamefont {Bozhko}, \citenamefont {Macêdo}, \citenamefont {Smith},
  \citenamefont {Holland}, \citenamefont {Danilin}, \citenamefont {Seferai},
  \citenamefont {Barbosa}, \citenamefont {Peroor}, \citenamefont {Goldman},
  \citenamefont {Nasti}, \citenamefont {Paul}, \citenamefont {Hadfield},
  \citenamefont {McVitie},\ and\ \citenamefont {Weides}}]{Baity2021}%
  \BibitemOpen
  \bibfield  {author} {\bibinfo {author} {\bibfnamefont {P.~G.}\ \bibnamefont
  {Baity}}, \bibinfo {author} {\bibfnamefont {D.~A.}\ \bibnamefont {Bozhko}},
  \bibinfo {author} {\bibfnamefont {R.}~\bibnamefont {Macêdo}}, \bibinfo
  {author} {\bibfnamefont {W.}~\bibnamefont {Smith}}, \bibinfo {author}
  {\bibfnamefont {R.~C.}\ \bibnamefont {Holland}}, \bibinfo {author}
  {\bibfnamefont {S.}~\bibnamefont {Danilin}}, \bibinfo {author} {\bibfnamefont
  {V.}~\bibnamefont {Seferai}}, \bibinfo {author} {\bibfnamefont
  {J.}~\bibnamefont {Barbosa}}, \bibinfo {author} {\bibfnamefont {R.~R.}\
  \bibnamefont {Peroor}}, \bibinfo {author} {\bibfnamefont {S.}~\bibnamefont
  {Goldman}}, \bibinfo {author} {\bibfnamefont {U.}~\bibnamefont {Nasti}},
  \bibinfo {author} {\bibfnamefont {J.}~\bibnamefont {Paul}}, \bibinfo {author}
  {\bibfnamefont {R.~H.}\ \bibnamefont {Hadfield}}, \bibinfo {author}
  {\bibfnamefont {S.}~\bibnamefont {McVitie}},\ and\ \bibinfo {author}
  {\bibfnamefont {M.}~\bibnamefont {Weides}},\ }\href
  {https://doi.org/10.1063/5.0054837} {\bibfield  {journal} {\bibinfo
  {journal} {Applied Physics Letters}\ }\textbf {\bibinfo {volume} {119}},\
  \bibinfo {pages} {033502} (\bibinfo {year} {2021})},\ \Eprint
  {https://arxiv.org/abs/https://doi.org/10.1063/5.0054837}
  {https://doi.org/10.1063/5.0054837} \BibitemShut {NoStop}%
\bibitem [{Note1()}]{Note1}%
  \BibitemOpen
  \bibinfo {note} {In our setup geometry, the magnetic field in the center of
  the cavity for both modes has the adequate perpendicular directions
  considering the volume of the sample.}\BibitemShut {Stop}%
\bibitem [{\citenamefont {Crescini}\ \emph {et~al.}(2020)\citenamefont
  {Crescini}, \citenamefont {Braggio}, \citenamefont {Carugno}, \citenamefont
  {Di~Vora}, \citenamefont {Ortolan},\ and\ \citenamefont
  {Ruoso}}]{crescini2020magnon}%
  \BibitemOpen
  \bibfield  {author} {\bibinfo {author} {\bibfnamefont {N.}~\bibnamefont
  {Crescini}}, \bibinfo {author} {\bibfnamefont {C.}~\bibnamefont {Braggio}},
  \bibinfo {author} {\bibfnamefont {G.}~\bibnamefont {Carugno}}, \bibinfo
  {author} {\bibfnamefont {R.}~\bibnamefont {Di~Vora}}, \bibinfo {author}
  {\bibfnamefont {A.}~\bibnamefont {Ortolan}},\ and\ \bibinfo {author}
  {\bibfnamefont {G.}~\bibnamefont {Ruoso}},\ }\href@noop {} {\bibfield
  {journal} {\bibinfo  {journal} {Communications Physics}\ }\textbf {\bibinfo
  {volume} {3}},\ \bibinfo {pages} {164} (\bibinfo {year} {2020})}\BibitemShut
  {NoStop}%
\bibitem [{\citenamefont {Spencer}\ \emph {et~al.}(1959)\citenamefont
  {Spencer}, \citenamefont {LeCraw},\ and\ \citenamefont
  {Clogston}}]{spencer1959}%
  \BibitemOpen
  \bibfield  {author} {\bibinfo {author} {\bibfnamefont {E.~G.}\ \bibnamefont
  {Spencer}}, \bibinfo {author} {\bibfnamefont {R.~C.}\ \bibnamefont
  {LeCraw}},\ and\ \bibinfo {author} {\bibfnamefont {A.~M.}\ \bibnamefont
  {Clogston}},\ }\href {https://doi.org/10.1103/PhysRevLett.3.32} {\bibfield
  {journal} {\bibinfo  {journal} {Phys. Rev. Lett.}\ }\textbf {\bibinfo
  {volume} {3}},\ \bibinfo {pages} {32} (\bibinfo {year} {1959})}\BibitemShut
  {NoStop}%
\bibitem [{\citenamefont {Suhl}(1957)}]{SUHL1957209}%
  \BibitemOpen
  \bibfield  {author} {\bibinfo {author} {\bibfnamefont {H.}~\bibnamefont
  {Suhl}},\ }\href
  {https://doi.org/https://doi.org/10.1016/0022-3697(57)90010-0} {\bibfield
  {journal} {\bibinfo  {journal} {Journal of Physics and Chemistry of Solids}\
  }\textbf {\bibinfo {volume} {1}},\ \bibinfo {pages} {209} (\bibinfo {year}
  {1957})}\BibitemShut {NoStop}%
\bibitem [{\citenamefont {Bryant}\ \emph {et~al.}(1988)\citenamefont {Bryant},
  \citenamefont {Jeffries},\ and\ \citenamefont {Nakamura}}]{bryant1988}%
  \BibitemOpen
  \bibfield  {author} {\bibinfo {author} {\bibfnamefont {P.}~\bibnamefont
  {Bryant}}, \bibinfo {author} {\bibfnamefont {C.}~\bibnamefont {Jeffries}},\
  and\ \bibinfo {author} {\bibfnamefont {K.}~\bibnamefont {Nakamura}},\ }\href
  {https://doi.org/10.1103/PhysRevLett.60.1185} {\bibfield  {journal} {\bibinfo
   {journal} {Phys. Rev. Lett.}\ }\textbf {\bibinfo {volume} {60}},\ \bibinfo
  {pages} {1185} (\bibinfo {year} {1988})}\BibitemShut {NoStop}%
\bibitem [{\citenamefont {Rezende}\ and\ \citenamefont
  {de~Aguiar}(1990)}]{rezende1990}%
  \BibitemOpen
  \bibfield  {author} {\bibinfo {author} {\bibfnamefont {S.}~\bibnamefont
  {Rezende}}\ and\ \bibinfo {author} {\bibfnamefont {F.}~\bibnamefont
  {de~Aguiar}},\ }\href {https://doi.org/10.1109/5.56906} {\bibfield  {journal}
  {\bibinfo  {journal} {Proceedings of the IEEE}\ }\textbf {\bibinfo {volume}
  {78}},\ \bibinfo {pages} {893} (\bibinfo {year} {1990})}\BibitemShut
  {NoStop}%
\bibitem [{\citenamefont {Lee}\ \emph {et~al.}(2023)\citenamefont {Lee},
  \citenamefont {Yamamoto}, \citenamefont {Umeda}, \citenamefont {Zollitsch},
  \citenamefont {Elyasi}, \citenamefont {Kikkawa}, \citenamefont {Saitoh},
  \citenamefont {Bauer},\ and\ \citenamefont {Kurebayashi}}]{kurebayashi2023}%
  \BibitemOpen
  \bibfield  {author} {\bibinfo {author} {\bibfnamefont {O.}~\bibnamefont
  {Lee}}, \bibinfo {author} {\bibfnamefont {K.}~\bibnamefont {Yamamoto}},
  \bibinfo {author} {\bibfnamefont {M.}~\bibnamefont {Umeda}}, \bibinfo
  {author} {\bibfnamefont {C.~W.}\ \bibnamefont {Zollitsch}}, \bibinfo {author}
  {\bibfnamefont {M.}~\bibnamefont {Elyasi}}, \bibinfo {author} {\bibfnamefont
  {T.}~\bibnamefont {Kikkawa}}, \bibinfo {author} {\bibfnamefont
  {E.}~\bibnamefont {Saitoh}}, \bibinfo {author} {\bibfnamefont {G.~E.~W.}\
  \bibnamefont {Bauer}},\ and\ \bibinfo {author} {\bibfnamefont
  {H.}~\bibnamefont {Kurebayashi}},\ }\href
  {https://doi.org/10.1103/PhysRevLett.130.046703} {\bibfield  {journal}
  {\bibinfo  {journal} {Phys. Rev. Lett.}\ }\textbf {\bibinfo {volume} {130}},\
  \bibinfo {pages} {046703} (\bibinfo {year} {2023})}\BibitemShut {NoStop}%
\bibitem [{\citenamefont {Hartwig}(1967)}]{hartwig_suppression_1967}%
  \BibitemOpen
  \bibfield  {author} {\bibinfo {author} {\bibfnamefont {C.~P.}\ \bibnamefont
  {Hartwig}},\ }\href {https://doi.org/10.1063/1.1709547} {\bibfield  {journal}
  {\bibinfo  {journal} {Journal of Applied Physics}\ }\textbf {\bibinfo
  {volume} {38}},\ \bibinfo {pages} {1220} (\bibinfo {year}
  {1967})}\BibitemShut {NoStop}%
\bibitem [{\citenamefont {Kobe}\ \emph {et~al.}(2017)\citenamefont {Kobe},
  \citenamefont {Chuma}, \citenamefont {Jamisola},\ and\ \citenamefont
  {Chose}}]{kobe_review_2017}%
  \BibitemOpen
  \bibfield  {author} {\bibinfo {author} {\bibfnamefont {O.~B.}\ \bibnamefont
  {Kobe}}, \bibinfo {author} {\bibfnamefont {J.}~\bibnamefont {Chuma}},
  \bibinfo {author} {\bibfnamefont {R.}~\bibnamefont {Jamisola}},\ and\
  \bibinfo {author} {\bibfnamefont {M.}~\bibnamefont {Chose}},\ }\href
  {https://doi.org/10.1016/j.jestch.2016.09.024} {\bibfield  {journal}
  {\bibinfo  {journal} {Engineering Science and Technology, an International
  Journal}\ }\textbf {\bibinfo {volume} {20}},\ \bibinfo {pages} {460}
  (\bibinfo {year} {2017})}\BibitemShut {NoStop}%
\bibitem [{\citenamefont {Cocconcelli}\ \emph {et~al.}(2024)\citenamefont
  {Cocconcelli}, \citenamefont {Tacchi}, \citenamefont {Erd\'elyi},
  \citenamefont {Maspero}, \citenamefont {Del~Giacco}, \citenamefont {Plaza},
  \citenamefont {Koplak}, \citenamefont {Cattoni}, \citenamefont {Silvani},
  \citenamefont {Madami}, \citenamefont {Papp}, \citenamefont {Csaba},
  \citenamefont {Kohl}, \citenamefont {Heinz}, \citenamefont {Pirro},\ and\
  \citenamefont {Bertacco}}]{cocconcelli2024}%
  \BibitemOpen
  \bibfield  {author} {\bibinfo {author} {\bibfnamefont {M.}~\bibnamefont
  {Cocconcelli}}, \bibinfo {author} {\bibfnamefont {S.}~\bibnamefont {Tacchi}},
  \bibinfo {author} {\bibfnamefont {R.}~\bibnamefont {Erd\'elyi}}, \bibinfo
  {author} {\bibfnamefont {F.}~\bibnamefont {Maspero}}, \bibinfo {author}
  {\bibfnamefont {A.}~\bibnamefont {Del~Giacco}}, \bibinfo {author}
  {\bibfnamefont {A.}~\bibnamefont {Plaza}}, \bibinfo {author} {\bibfnamefont
  {O.}~\bibnamefont {Koplak}}, \bibinfo {author} {\bibfnamefont
  {A.}~\bibnamefont {Cattoni}}, \bibinfo {author} {\bibfnamefont
  {R.}~\bibnamefont {Silvani}}, \bibinfo {author} {\bibfnamefont
  {M.}~\bibnamefont {Madami}}, \bibinfo {author} {\bibfnamefont
  {A.}~\bibnamefont {Papp}}, \bibinfo {author} {\bibfnamefont {G.}~\bibnamefont
  {Csaba}}, \bibinfo {author} {\bibfnamefont {F.}~\bibnamefont {Kohl}},
  \bibinfo {author} {\bibfnamefont {B.}~\bibnamefont {Heinz}}, \bibinfo
  {author} {\bibfnamefont {P.}~\bibnamefont {Pirro}},\ and\ \bibinfo {author}
  {\bibfnamefont {R.}~\bibnamefont {Bertacco}},\ }\href
  {https://doi.org/10.1103/PhysRevApplied.22.064063} {\bibfield  {journal}
  {\bibinfo  {journal} {Phys. Rev. Appl.}\ }\textbf {\bibinfo {volume} {22}},\
  \bibinfo {pages} {064063} (\bibinfo {year} {2024})}\BibitemShut {NoStop}%
\end{thebibliography}%

\end{document}